\documentclass[amsmath, amsfonts, superscriptaddress, showpacs, twocolumn, prb]{revtex4-1}
\usepackage{graphicx}
\usepackage{epsfig}
\usepackage{bm}
\usepackage{dcolumn}
\usepackage{amsmath}
\usepackage{amssymb}

\begin{document}

\title{Boltzmann-Langevin theory of Coulomb drag}

\author{W.~Chen}
\affiliation{Institute for Advanced Study, Tsinghua University, Beijing, 100084, China}

\author{A.~V.~Andreev}
\affiliation{Department of Physics, University of Washington,
Seattle, Washington 98195, USA}

\author{A.~Levchenko}
\affiliation{Department of Physics and Astronomy, Michigan State
University, East Lansing, Michigan 48824, USA}
\affiliation{Department of Physics, University of Wisconsin-Madison, Madison, Wisconsin 53706, USA}

\begin{abstract}
We develop a Boltzmann-Langevin description of Coulomb drag effect in clean double-layer systems with large interlayer separation $d$ as compared to the average interelectron distance $\lambda_F$. Coulomb drag arises from density fluctuations with spatial scales of order $d$. At low temperatures, their characteristic frequencies exceed the intralayer equilibration rate of the electron liquid, and Coulomb drag may be treated in the collisionless approximation. As temperature is raised, the electron mean free path becomes short due to electron-electron scattering. This leads to local equilibration of electron liquid, and consequently drag is determined by hydrodynamic density modes. Our theory applies to both the collisionless and the hydrodynamic regimes, and it enables us to describe the crossover between them. We find that drag resistivity exhibits a nonmonotonic temperature dependence with multiple crossovers at distinct energy scales.  At the lowest temperatures, Coulomb drag is dominated by the particle-hole continuum, whereas at higher temperatures of the collision-dominated regime it is governed by the plasmon modes. We observe that fast intralayer equilibration mediated by electron-electron collisions ultimately renders a stronger drag effect.
\end{abstract}

\date{June 2, 2015}

\pacs{71.27.+a, 72.10.-d, 73.40.Ei, 73.63.Hs}

\maketitle

\section{Introduction}

Since the pioneering papers by Pogrebinskii~\cite{Pogrebinskii} and Price,~\cite{Price} their idea of Coulomb drag has evolved into the independent and fruitful field of research in condensed-matter physics. The initial motivation was to propose an experiment that would provide a direct measure of the rate of electron-electron collisions not masked by other competing relaxation channels, such as electron-impurity or electron-phonon collisions. This became possible and was realized in the electrically isolated but interactively coupled double-layer systems.~\cite{Solomon1,Solomon2,Gramila,Sivan,Eisenstein} When current is driven through one layer while the voltage across the other layer is measured, the resulting nonlocal drag resistivity is a direct probe of the rate of momentum transfer between the two layers via electron-electron scattering. In many practical cases,  Coulomb drag measurements provide incredibly sensitive tools for revealing electronic correlations, which are not readily accessible via more standard transport experiments in single-layer samples. Recent results on drag resistivity in double-layer heterostructures made of graphene~\cite{Tutuc,Geim,Kim} triggered a flood of theoretical work~\cite{Katsnelson,Hwang,Peres,Narozhny-1,Scharf,Guinea,Polini,Fritz} including new proposals for the mechanism of this phenomenon in the case of tightly nested layers.~\cite{Song-1,Song-2,Song-3,Schutt,Titov,Narozhny-2}

Coulomb drag resistivity $\rho_D$ is extremely sensitive to temperature $T$, magnetic field $B$, interlayer spacing $d$, intralayer density $n$ or density mismatch between the layers, and intralayer mean free path $\ell$, which can be dominated by either impurity scattering in the disordered case or by electron-electron collisions in clean systems. The strength of correlation effects can be conveniently described by the electron gas parameter $r_s=(\pi na^2_B)^{-1/2}$, where $a_B=\epsilon/me^2$ is the effective Bohr radius in the host material of a quantum well with dielectric constant $\epsilon$.

In initial measurements~\cite{Solomon1,Solomon2,Gramila,Sivan,Eisenstein} both the characteristic dependence of the drag resistivity on the various parameters and the magnitude of the effect were in very good agreement with theoretical predictions.~\cite{Laikhtman,Solomon1,Solomon2,Maslov,Smith,MacDonald,Kamenev,Flensberg,Rojo} In particular, at relatively low temperatures and in the clean limit,
\begin{equation}\label{r-D-ph}
\frac{\rho_D(T)}{\rho_Q}\simeq\left(\frac{T}{E_F}\right)^2\left(\frac{1}{\kappa d}\right)^2\left(\frac{1}{k_Fd}\right)^2,
\end{equation}
where $k_F$ and $E_F$ are Fermi momentum and energy, respectively, $\kappa\simeq r_sk_F$ is the inverse Thomas-Fermi screening radius, and $\rho_Q=2\pi/e^2$ is quantum of resistance (throughout the paper, $\hbar=k_B=1$). These early experiments were performed on quantum wells with the typical electron density $n\sim10^{11}$cm$^{-2}$ and layer separation $d\sim 250\AA$. That translates into $E_F\sim 60$K, $k_Fd\gg1$ and corresponds to the weakly interacting limit $r_s\sim 1$, which was explored in theoretical calculations.

Subsequent detailed experimental investigations of drag resistivity posed theoretical challenges. For example, measurements~\cite{Gramila-2,Kellogg} on samples with densities and mobilities comparable to that of early measurements, but with much higher interlayer separations (up to $d\sim5000\AA$), revealed that $\rho_D/T^2$ ceases to be  a constant but acquires a peculiar temperature dependence, while the overall magnitude of the drag resistivity significantly exceeds the expected value. Such a dramatic disagreement between experimental findings and the theoretical predictions of Eq.~\eqref{r-D-ph} was attributed to an additional contribution to drag effect due to virtual-phonon exchange.~\cite{MacDonald-phonons} It was also observed that at higher temperatures, drag resistivity becomes a nonmonotonic function of $T$. It exhibits a well-pronounced peak at $T\sim E_F$ followed by a rapid decay at higher temperatures.~\cite{Hill,Noh} The difficulty of explaining this feature within the phonon drag mechanism prompted consideration of the plasmon contribution to Coulomb drag,~\cite{Flensberg-plasmons-1,Flensberg-plasmons-2} which yields an enhancement of drag at temperatures of the order of the characteristic plasmon energies. Further challenges to the theory of Coulomb drag were posed by experiments in samples with low carrier density, $n\sim 10^9$cm$^{-2}$, where inter-electron interactions are strong, $r_s\gg1$.~\cite{Pillarisetty-1,Pillarisetty-2}  In such double-layers, even at low temperatures, $T\ll E_F$, the drag resistance is one to two orders of magnitude larger than expected on the basis of a simple extrapolation of the small $r_s$ results. Furthermore, the power exponent in the temperature dependence deviates from being simply quadratic, drag has unexpected behavior on the density mismatch, and the system has an anomalous response to a magnetic field.

For electron liquids with $r_s\gg 1$ there exists a wide  temperature interval, $E_F<T< r_s E_F$, in which the liquid is strongly correlated but is not quantum degenerate, and hence it may not be described by the Landau Fermi-liquid theory. Within this interval, one can further distinguish between the classical,  $\omega_D <T <r_s E_F$, and semiquantum,~\cite{Andreev_semiquantum,Spivak_review} $E_F <T< \omega_D$, regimes (here $\omega_D \sim E_F \sqrt{r_s}$ is the Debye frequency in the liquid). Theory of Coulomb drag in this temperature interval has not been developed. A detailed consideration of Coulomb drag based on an extrapolation of Fermi-liquid-based formulas to the region where $r_s>1$ has been carried out in Ref.~\onlinecite{Stern} in an attempt to address the data of Ref.~\onlinecite{Pillarisetty-1}, however such analysis can be qualitatively justified only for the temperature range $T<E_F/(k_Fd)$. Another approach to the theory of Coulomb drag in systems with $r_s\gg 1$ was developed in the hydrodynamic regime,~\cite{Hydro-drag} where the mean free path due to intralayer electron-electron collisions becomes shorter than other relevant length scales. In this case, the drag resistivity can be expressed in terms of the viscosity and thermal conductivity of the electron liquid. The hydrodynamic approach does not assume the Fermi-liquid behavior of the electron fluid, but it also applies to classical and semiquantum strongly correlated liquids for which experimental realization of the hydrodynamic regime is more realistic. As the temperature is lowered, the Fermi-liquid theory becomes applicable. At the same time, the equilibration length due to intralayer electron-electron collisions becomes longer and the system eventually crosses over into the collisionless regime of Coulomb drag.

Previous microscopic calculations of Coulomb drag were made under a tacit assumption of the collisionless regime with respect to intralayer electron-electron scattering, namely when the intralayer mean free path exceeds interlayer separation $\ell\gg d$. The only exception is the paper Ref.~\onlinecite{Polyakov} where crucial importance of the intralayer equilibration has been emphasized in the context of drag between one-dimensional quantum wires. In the two-dimensional case, at large interlayer spacings, the crossover from the collisionless to the hydrodynamic regime occurs within the range of applicability of the Fermi-liquid theory. This enables construction of a microscopic description of Coulomb drag in the entire crossover interval between the collisionless and the hydrodynamic regimes.

In the present work, we develop a theory of Coulomb drag for an arbitrary relation between the interlayer spacing $d$ and the intralayer equilibration length $\ell$. We find that the crossover from the collisionless, $\ell\gg d$, to the collision-dominated (hydrodynamic) regime, $\ell\ll d$, is hierarchical and is characterized by several parametrically distinct energy scales. Depending on the temperature, drag is dominated either by the particle-hole continuum or plasmon modes. In addition to reproducing the previously known $\propto T^2$ temperature dependence in the collisionless regime [Eq.~(\ref{r-D-ph})], we clarify the more subtle linear in $\propto T$ term briefly discussed in Ref.~\onlinecite{Smith}. More importantly, we identify a regime in which the drag resistivity is dominated by plasmons and follows a $T^3$ temperature dependence. Apart from being relevant to experiments, our work provides an alternative approach to the Coulomb drag effect that is based on the stochastic Boltzmann-Langevin kinetic equation.~\cite{Kadomtsev,Kogan-Shulman}

The paper is organized as follows.  In Sec.~\ref{Sec-Qual} we present a qualitative discussion for the Coulomb drag effect in the crossover region from the collisionless to the hydrodynamic regime. In Sec. \ref{Sec-Formalism} we derive a general formula for drag resistivity using the formalism of the stochastic kinetic equation. In Sec. \ref{sec:boost} we focus on the situation in which the interlayer momentum relaxation rate due to drag, $1/\tau_D = ne^2 \rho_D/m$, is smaller than the equilibration rate of the electron liquid, $1/\tau \sim r^2_sT^2/E_F$, when the description of drag is greatly simplified. This situation is realized at sufficiently large intralayer separations. In this regime, the electron liquid in the layers may be assumed to be in local equilibrium characterized by a drift velocity in each layer. To obtain specific predictions for the temperature dependence of the drag resistivity in Sec.~\ref{sec:tau_approximation}, we adopt a model collision integral characterized by a single relaxation rate $1/\tau$. In Sec. \ref{Sec-Results} we analyze our result in various temperature regimes, and we uncover the structure of the crossover in temperature dependence of the drag resistivity. We find that at low temperatures, drag is dominated by the particle-hole continuum, whereas in the collision-dominated regime of higher temperatures, drag is governed by the plasmon modes. The ratio of $\rho_D(T)/T^2$ is a nonmonotonic function of temperature that exhibits a broad peak. In Sec. \ref{Sec-Discussion} we discuss our findings and compare them with available experimental results. In our summary, we also discuss open questions and directions for future research related to magnetodrag phenomena.

\section{Qualitative discussion}\label{Sec-Qual}

Coulomb drag originates from interaction between fluctuations of electron densities in the two layers with spatial scales of the order of the interlayer distance $d$. At large interlayer separations, $k_F d\gg 1$, such fluctuations involve a large number of particles. To leading order in $1/(k_F d)$, the instantaneous electron density at spatial scales of order $d$ is given by the mean density of the electron fluid. In this approximation, however, the Coulomb drag vanishes. The fluctuations of electron density that result in drag ultimately arise from the discrete nature of charge carriers, and at small spatial scales they have the character of ``shot noise''.   At spatial scales of order $d$, the characteristic frequencies and character of propagation of density fluctuations that govern drag, depend sensitively on the mean free time of the electrons due to intralayer electron-electron scattering, $\tau$.

If the characteristic frequencies of density fluctuations responsible for Coulomb drag exceed $1/\tau$, the influence of intralayer electron-electron scattering on the dynamics of such fluctuations may be neglected. We refer to this regime as collisionless. We stress that even in the collisionless drag regime, electron-electron scattering is essential for establishing the steady state. Only its influence on the propagation of density fluctuations is negligible. In the opposite case, i.e., the collision-dominated drag regime, electron-electron collisions play a crucial role not only in establishing the character of steady-state flow, but also in the propagation of density fluctuations responsible for drag.

In the collisionless regime,  the density fluctuations consist of ballistically propagating particle-hole excitations and of plasmons. The characteristic energy scales associated with these two types of density fluctuations are, respectively,
\begin{equation}\label{eq:scales}
 T_d=\frac{E_F}{k_F d}, \qquad \omega_{pl}=T_d \sqrt{\kappa d}.
\end{equation}
At $T<T_d$, drag is dominated by the particle-hole continuum and follows the quadratic temperature dependence, Eq.~(\ref{r-D-ph}). The $T^2$ scaling can be simply understood from the phase-space argument for the low-temperature two-particle scattering arising from the $\sim T$ smearing of both the initial and final states near the Fermi surface. The power law $\propto d^{-4}$ falloff of $\rho_D$ follows from the screening properties of the Coulomb potential.

At higher temperatures, $T>T_d$, drag continues to be dominated by the particle-hole continuum but  the quadratic temperature dependence is replaced by the linear one, $\rho_D\propto T/T_d$,~\cite{Smith} because of phase-space limitations associated with the predominance of small-angle scattering. The crossover to the collision-dominated regime for the particle-hole excitations happens at $1/\tau \sim T_d$. For Fermi-liquids this occurs at $T_c\sim \sqrt{E_F T_d}\sim\omega_{pl}$. The crossover to the collision-dominated regime for plasmons, $1/\tau > \omega_{pl}$, occurs at a much higher temperature scale $T_h\sim \sqrt{E_F \omega_{pl}}$. Above this temperature the electron liquid enters the hydrodynamic regime where the distinction between the particle-hole continuum and plasmons is no longer meaningful. Drag resistivity in the hydrodynamic regime follows\cite{Hydro-drag} the $1/T$ behavior. The crossover regime
\begin{equation}
\omega_{pl}<T< T_h
\end{equation}
has not been previously explored. Note that the extrapolation of the hydrodynamic result~\cite{Hydro-drag}
\begin{equation}\label{r-D-h}
\frac{\rho_D(T)}{\rho_Q}\simeq \frac{E_F}{T}\left(\frac{1}{k_Fd}\right)^5
\end{equation}
to temperatures $T\sim T_h$ yields a greater drag resistivity than the extrapolation of the ballistic result to $T_h$ by a parametrically large factor $k_Fd\gg1$. One concludes then that fast intralayer equilibration mediated by electron-electron collisions ultimately renders stronger drag effect. The detailed comparison of various regimes is given in Sec.~\ref{Sec-Results}.

To get some physical insight into the origin of the apparent mismatch between collisionless and hydrodynamic answers for the drag resistivity, one should recall that hydrodynamic description follows from a more general kinetic theory by retaining only zero-modes of the collision integral. These modes correspond to the conservation laws of particle number, momentum, and energy of the liquid, and they are described by the hydrodynamic equations. The more general kinetic description captures not only the hydrodynamic modes but also modes that are relaxed by electron-electron scattering. Their fluctuations decay rapidly with the typical rate of inelastic collisions and govern the crossover regime.

Below we develop a general theory of Coulomb drag in clean double layers with large interlayer spacing that is valid in both collisionless and collision-dominated regimes, and it enables us to explore the crossover between them. The dynamics of density fluctuations can be described using the generalization of the Boltzmann-Langevin formalism~\cite{Kadomtsev,Kogan-Shulman} to nonideal gases.~\cite{Klimontovich} In this approach, one describes the state of the system by a fluctuating distribution function $f=\bar f+\delta f$ averaged over a physically microscopic spatial scales (of order $d$ in our case) containing a large number of particles. Time evolution is described by the system of equations for the average distribution function $\bar f$ and  the fluctuations $\delta f$. The evolution equation for the fluctuating part $\delta f$ is the Boltzmann equation with a fluctuating Langevin source whose variance is determined by the average distribution function $\bar f$. The short-range part of the Coulomb interaction between electrons is described by the collision integral whereas its long-range part enters the Boltzmann equation in the form of an external potential related to the density of fluctuation by the Poisson equation.  The evolution equation for the average distribution function $\bar f$ also differs from the standard Boltzmann equation. The difference arises because in contrast to the assumption of complete relaxation of correlations that underlies the Boltzmann equation, only relaxation of short-range correlations is assumed, whereas long-range correlations are taken into account.~\cite{Klimontovich} As a result, the evolution equation for the average distribution function $\bar f$ contains the correlator of long-range density fluctuations of the distribution functions.

\section{Boltzmann-Langevin approach to Coulomb drag}\label{Sec-Formalism}

There are three different computational approaches to the problem of Coulomb drag (CD) resistivity, which are based on either the Kubo formula,~\cite{Kamenev,Flensberg} and closely related to it the memory-function formalism,~\cite{MacDonald,Rojo} or alternatively the kinetic equation.~\cite{Smith,Flensberg-plasmons-1,Flensberg-plasmons-2} The latter approach has certain advantages over the other methods since it allows one to account on equal footing for inter- and intralayer interactions, and in principle it can be generalized to nonequilibrium situations. It should be noted that previous treatments of the CD problem based on the kinetic equation~\cite{Smith,Flensberg-plasmons-1,Flensberg-plasmons-2} had been carried out in the main kinetic approximation, namely without inclusion of the stochastic Langevin forces. Furthermore, the kinetic equation for the electron distribution function due to intralayer collisions has not been solved explicitly. Instead, the distribution function was extracted from the argument of the Galilean boost in the moving frame with electron liquid and the electron polarization function was then computed for the bare noninteracting limit. Because of these reasons, such a description does not adequately capture the crossover to the collision-dominated regime. We will show that not only is the nonequilibrium part of the electron distribution function important in the calculation of the drag response, but there are also corresponding corrections to the electron polarization function that give rise to contributions to drag resistivity. Moreover, the functional form of the polarization function is strongly affected by intralayer collisions, which thus require separate serious considerations. This physical picture has certain parallels with the problem of nonequilibrium fluctuations and the shot noise. Indeed, it is known that electron collisions strongly modify the spectral function of current fluctuations and change the Fano factor.~\cite{Nagaev} Drag may be interpreted as the rectification effect of nonequilibrium fluctuations of electron liquid~\cite{Drag-QPC} so that one naturally expects that equilibration processes play a prominent role for this phenomenon. This observation suggests using Boltzmann-Langevin kinetic theory, which was extremely fruitful in the context of the shot-noise problem, in application to the drag effect.

\subsection{Boltzmann-Langevin kinetic equation}\label{sec:Boltzmann-Langevin}

We consider a Coulomb drag double-layer setup in which the active drive-layer is driven out of equilibrium by an applied bias while the induced response is measured in the other passive drag layer. We describe the state of the electron liquid within each of the layers labeled by an index $\imath=\uparrow,\downarrow$ by the respective distribution functions $f^\imath=\bar{f}^\imath+\delta f^\imath$, where $\bar{f}^\imath$ is the time average distribution function and $\delta f^\imath$ is its fluctuating part. The general Boltzmann-Langevin (BL) kinetic equation is~\cite{Kogan-Shulman}
\begin{equation}\label{KE}
[\partial_t+\mathbf{v}\cdot\bm{\nabla}-e(\mathbf{E}^\imath-\bm{\nabla}\phi^\imath)\partial_{\mathbf{p}}]f^\imath=
\mathrm{St}\{f^\imath\}+\delta J^\imath.
\end{equation}
The right-hand side of this equation describes the flux of particles into a given phase-space point due to collisions. The average flux is described by the collision integral $\mathrm{St}\{f^\imath\}$. The stochastic nature of collisions causes fluctuations in the probability flux, which are described by the extraneous source $\delta J^\imath$. The second-order correlation function of the extraneous sources is local, $\langle \delta J(\mathbf{p},\mathbf{r},t)\delta J(\mathbf{p}',\mathbf{r}',t')\rangle \propto \delta(t-t') \delta(\mathbf{r}-\mathbf{r}')$, reflecting the locality of collisions, and it has been evaluated by Kogan and Shulman~\cite{Kogan-Shulman} for various types of scattering processes. Its specific form depends on the form of the collision integral, and it can be found in Refs.~\onlinecite{Kogan-Shulman,Nagaev}.

The fluctuating electric potentials $\phi^\imath$ in Eq.~(\ref{KE}) are related to the distribution functions in both layers by the Poisson equation
\begin{equation}\label{Poisson-phi}
\phi^\imath_{\omega,q}=-\frac{2\pi e}{\epsilon q}\int d\Gamma_p[\delta f^\imath_{\omega,q}(\mathbf{p})+e^{-qd}\delta f^{-\imath}_{\omega,q}(\mathbf{p})],
\end{equation}
where $d\Gamma=2 d^2p/(2\pi)^2$ denotes the density of states in two-dimensional momentum space with the factor of $2$ accounting for the spin. In addition, in Eq.~(\ref{Poisson-phi}) we also used the Fourier representation, and we denoted the interlayer distance by $d$ and $q=|\mathbf{q}|$. Isolating the fluctuating part of Eq.~(\ref{KE}), we obtain in the Fourier representation
\begin{equation}\label{KE-linearized}
(-i\omega+i\mathbf{v}\mathbf{q}-e\mathbf{E}^\imath\partial_{\mathbf{p}})\delta f^\imath_{\omega,q}
+ie\mathbf{q}\phi^\imath_{\omega,q}\partial_{\mathbf{p}}\bar{f}^\imath\!=\!
\delta\mathrm{St}\{f^\imath\}+\delta J^\imath_{\omega,q},
\end{equation}
where $\delta\mathrm{St}\{f^\imath\}$ is the fluctuating part of the collision integral. To leading order in the fluctuations, the latter may be obtained by expanding the collision integral to the linear order in $\delta f^\imath$ around $\bar{f}^\imath$. For the average part of the distribution function, we get from Eq.~(\ref{KE})
\begin{equation}\label{KE-average}
-e\mathbf{E}^\imath\partial_{\mathbf{p}}\bar{f}^\imath+e\langle\bm{\nabla}\phi^\imath\partial_{\mathbf{p}}\delta f^\imath\rangle=\mathrm{St}\{\bar{f}^\imath\},
\end{equation}
where angular brackets $\langle\cdots\rangle$ denote averaging over the fluctuations.

Equations (\ref{Poisson-phi}), (\ref{KE-linearized}), and (\ref{KE-average}) supplemented by the expression for the variance of the Langevin fluxes $\delta J^\imath$ in terms of the average distribution function $\bar{f}^\imath$ constitute a closed system describing the fluctuations and Coulomb drag in the double-layer setup.  Drag resistivity $\rho_D=\mathbf{E}^\uparrow/\mathbf{j}^\downarrow$ is defined as the ratio between the electric field generated in the drag layer in response to the current flow in the drive layer.

At this point, we briefly outline the program of calculations that will lead us to the general expression for $\rho_D$. First, we need to find the Green's function of the Boltzmann equation (\ref{KE-linearized}). Next we will use the Boltzmann-Langevin scheme to evaluate the density-density correlations and consequently the dragging force. To perform averaging over the stochastic fluxes, we will assume that correlators of Langevin sources have equilibrium form in each layer, one moving and one at rest, thus employing the fluctuation-dissipation relation in the boosted reference frame.

\subsection{Drag force}

Using the general approach summarized above, the drag resistivity may be determined by solving for the nonequilibrium electron distribution function that arises in response to a driving (staggered in the layer index) electric field and evaluating the resulting staggered current. Alternatively, it may be advantageous to evaluate the interlayer momentum transfer rate (drag force)  for a given nonequilibrium distribution of electrons. In the present section, we derive a general expression for the drag force in terms of the electron distribution function in the layers. In Sec.~\ref{sec:boost} we use this expression to evaluate drag resistivity at large interlayer separations.

A formal solution of Eq.~(\ref{KE-linearized}) may be written in the form
\begin{equation}\label{KE-sol}
\delta f^\imath_{\omega,q}=\hat{K}^\imath_{\omega,q}(-ie\mathbf{q}\phi^\imath_{\omega,q}\partial_{\mathbf{p}}\bar{f}^\imath+\delta J^\imath_{\omega,q}),
\end{equation}
where
\begin{equation}
\left(\hat{K}^\imath_{\omega,q}\right)^{-1}\delta f^\imath_{\omega,q}=
(-i\omega+i\mathbf{v}\mathbf{q}-e\mathbf{E}^\imath\partial_{\mathbf{p}})\delta f^\imath_{\omega,q}-\delta\mathrm{St}\{f^\imath\}
\end{equation}
is the resolvent of the Boltzmann equation. We note that if the stationary distribution function $\bar{f}^\imath$ does not correspond to equilibrium, the linearized collision integral does not generally have zero modes. However, if deviations from equilibrium are small, some of the eigenvalues of the linearized collision integral are expected to be anomalously small; the latter correspond to hydrodynamic modes. We also note that the source on the right-hand side of the above equation is localized in phase space in the vicinity of the Fermi surface.

Multiplying Eq.~(\ref{KE-average}) by $\mathbf{p}$ and integrating by parts, we obtain
\begin{equation}\label{D-Force-Def}
e\mathbf{E}^\imath\bar{n}^\imath=e\langle\bm{\nabla}\phi^\imath\delta n^\imath\rangle,
\end{equation}
where we used conservation of momentum in collisions $\int d\Gamma\mathbf{p}\mathrm{St}\{\bar{f}\}$, and we introduced the average density $\bar{n}^\imath=\int d\Gamma\bar{f}^\imath$ in layer-$\imath$ and its fluctuations $\delta n^\imath=\int d\Gamma\delta f^\imath$.

Using the Poisson equation (\ref{Poisson-phi}) we can reduce Eq.~(\ref{D-Force-Def}) to
\begin{equation}
e\mathbf{E}^\imath\bar{n}^\imath=-2\pi e^2\sum_{q,\omega}\frac{i\mathbf{q}e^{-qd}}{\epsilon q}\left(\delta\vec{n}\otimes\delta\vec{n}^T\right)^{\imath,-\imath}_{\omega,q}.
\end{equation}
Here we wrote density fluctuations in the two layers in the form of a column vector and introduced the spectra power of density fluctuations in the standard way
\begin{equation}
\langle\delta\vec{n}_{\omega,q}\otimes\delta\vec{n}^T_{\omega',q'}\rangle=(2\pi)^3\delta(\omega+\omega')\delta(\mathbf{q}+\mathbf{q}')\left(\delta\vec{n}\otimes\delta\vec{n}^T\right)_{\omega,q}.
\end{equation}
Summation over the frequency and momenta implies $\sum_{q,\omega}=\int\frac{d\omega}{2\pi}\int d\Gamma_q$. Changing the variables $(\omega,\mathbf{q})\to(-\omega,-\mathbf{q})$ under the integral we can rewrite the drag force as
\begin{equation}\label{D-Force-Trace}
e\mathbf{E}^\imath\bar{n}^\imath=-\pi e^2\sum_{q,\omega}\frac{\mathbf{q}e^{-qd}}{\epsilon q}\mathrm{Tr}\left[\hat{\sigma}_y\left(\delta\vec{n}\otimes\delta\vec{n}^T\right)_{\omega,q}\right],
\end{equation}
where we denote the Pauli matrices in the layer space by $\hat{\sigma}$. From Eq.~(\ref{KE-sol}) we get
\begin{equation}\label{Phi-Pi}
\delta n^\imath_{\omega,q}=\delta n^{\imath,ext}_{\omega,q}-e\phi^\imath_{\omega,q}\Pi^\imath_{\omega,q}.
\end{equation}
Here $\delta n^{\imath,ext}_{\omega,q}$ denotes the extrinsic density fluctuations that would be induced by Langevin sources in the Fermi gas in the absence of long-range Coulomb interactions
\begin{equation}
\delta n^{\imath,ext}_{\omega,q}=\int d\Gamma_pd\Gamma_kK^\imath_{\omega,q}(\mathbf{p},\mathbf{k})\delta J^\imath_{\omega,q}(\mathbf{k}).
\end{equation}
The polarization operator function
\begin{equation}
\Pi^\imath_{\omega,q}=\int d\Gamma_pd\Gamma_kK^\imath_{\omega,q}(\mathbf{p},\mathbf{k})i\mathbf{q}\partial_{\mathbf{p}}\bar{f}^\imath
\end{equation}
describes the response to a fluctuating potential in a nonequilibrium steady state $\bar{f}^\imath(\mathbf{p})$ and also in the presence of the electric field $\mathbf{E}^\imath$. We can rewrite Eq.~(\ref{Phi-Pi}) in the form
\begin{equation}\label{Pi-n}
(\hat{1}-\hat{\Pi}_{\omega,q}\hat{V}_q)\delta\vec{n}_{\omega,q}=\delta\vec{n}^{ext}_{\omega,q},
\end{equation}
where matrices (in the layer index) of polarization operator and interaction potential are denoted as follows:
\begin{equation}\label{Pi-V}
\hat{\Pi}_{\omega,q}=\delta_{\imath\imath'}\Pi^\imath_{\omega,q},\quad \hat{V}_q=\frac{2\pi e^2}{\epsilon q}\left(\begin{array}{cc}1 & e^{-qd} \\ e^{-qd} & 1 \end{array}\right).
\end{equation}
From the last two equations, we read off the power spectrum of density fluctuations in the form
\begin{eqnarray}
&&\hspace{-.35cm}(\delta\vec{n}\otimes\delta\vec{n}^T)_{\omega,q}=\nonumber\\
&&\hspace{-.35cm}(1-\hat{\Pi}_{\omega,q}\hat{V}_q)^{-1}\hat{N}_{\omega,q}(1-\hat{V}_q\hat{\Pi}_{-\omega,-q})^{-1},
\end{eqnarray}
where $\hat{N}_{\omega,q}=(\delta\vec{n}^{ext}\otimes\delta\vec{n}^{ext, T})_{\omega,q}$. Substituting this expression into the trace in Eq.~(\ref{D-Force-Trace}), we get
\begin{eqnarray}
&&\mathrm{Tr}\!\left[\hat{\sigma}_y\left(\delta\vec{n}\otimes\delta\vec{n}^T\right)_{\omega,q}\right]\!=\mathrm{Tr}\!\left[\hat{V}^{-1}_q\hat{\sigma}_y\hat{V}^{-1}_q\right.\nonumber\\
&&\left.(\hat{V}^{-1}_q-\hat{\Pi}_{\omega,q})^{-1}\hat{N}_{\omega,q}
(\hat{V}^{-1}_q-\hat{\Pi}_{-\omega,-q})^{-1}\right].
\end{eqnarray}
By employing Eq.~(\ref{Pi-V}), we notice that
\begin{equation}
\hat{V}^{-1}_q\hat{\sigma}_y\hat{V}^{-1}_q=\left(\frac{\epsilon q}{2\pi e^2}\right)^2\frac{\hat{\sigma}_y}{1-e^{-2qd}}
\end{equation}
and therefore the formula for the dragging force reduces to
\begin{eqnarray}\label{D-Force-Trace-v-P}
&&\hspace{-.15cm}e\mathbf{E}^i\bar{n}^i=-\frac{\pi e^2}{2\epsilon \kappa^2}\sum_{q,\omega}\frac{q^2\mathbf{e_q}}{\sinh(qd)}
\nonumber\\
&&\hspace{-.15cm}\mathrm{Tr}\!\left[\hat{\sigma}_y\!\left(\frac{q}{\kappa}\hat{v}^{-1}_q-\hat{P}_{\omega,q}\right)^{\!-1}\!\!\hat{N}_{\omega,q}\!
\left(\frac{q}{\kappa}\hat{v}^{\!-1}_q-\hat{P}_{-\omega,-q}\right)^{-1}\right]
\end{eqnarray}
where $\mathbf{e_q}$ is the unit vector in the direction of $\mathbf{q}$. Here we also introduced the dimensionless interaction matrix $\hat{v}_q=(\epsilon q/2\pi e^2)\hat{V}_q$, the dimensionless polarization operator $\hat{P}=\nu^{-1}\hat{\Pi}$, where $\nu$ is the single-particle density of states, and the inverse Thomas-Fermi screening radius $\kappa=2\pi\nu e^2/\epsilon$. The above equation expresses the drag force in terms of the density response functions in each layer, and the correlator of extrinsic density fluctuations. We note here that the latter in general can not be reduced to a density response function. In equilibrium, $\hat{N}_{\omega,q}$ and $\hat{P}_{\omega,q}$ are proportional to the unity matrix, and the interaction matrix $\hat{v}_q$ has only $\hat{\sigma}_0$ and $\hat{\sigma}_x$ components. As a consequence, the trace in the integrand of Eq.~(\ref{D-Force-Trace-v-P}) vanishes resulting in zero drag force, as it should be.

For the further convenience with the intermediate steps of calculation we introduce the following matrices
\begin{subequations}
\begin{equation}\label{P}
\hat{\mathcal{P}}=\frac{qe^{qd}}{2\kappa\sinh(qd)}\hat{\sigma}_0-\hat{P}_{\omega,q},\\
\end{equation}
\begin{equation}
\hat{\mathcal{P}}^*=\frac{qe^{qd}}{2\kappa\sinh(qd)}\hat{\sigma}_0-\hat{P}_{-\omega,-q},\\
\end{equation}
\begin{equation}\label{V}
\hat{\mathcal{V}}=\frac{q}{2\kappa\sinh(qd)}\hat{\sigma}_x.
\end{equation}
\end{subequations}
We remind the reader that the polarization operator matrix is diagonal in the layer index, however it is proportional to $\hat{\sigma}_0$ only for the case of identical layers. We will assume this case for the simplicity of further considerations, and we will provide the necessary generalization for the case of unequal layers at the end of this section. With these new notations we observe that $\left(\frac{q}{\kappa}\hat{v}^{-1}_q-\hat{P}_{\omega,q}\right)^{-1}=(\hat{\mathcal{P}}-\hat{\mathcal{V}})^{-1}$ and $\left(\frac{q}{\kappa}\hat{v}^{-1}_q-\hat{P}_{-\omega,-q}\right)^{-1}=(\hat{\mathcal{P}}^*-\hat{\mathcal{V}})^{-1}$. Next, let us denote nonequilibrium corrections to various quantities by $\Delta$ so that to the linear order in those we get from (\ref{D-Force-Trace-v-P})
\begin{equation}
\label{Long-Trace}
\begin{split}
e\mathbf{E}^i\bar{n}^i=-\frac{\pi e^2}{2\epsilon\kappa^2}\sum_{q,\omega}\frac{q^2\mathbf{e_q}}{
\sinh(qd)}\mathrm{Tr}\left[(\hat{\mathcal{P}}^*-\hat{\mathcal{V}})^{-1}
\hat{\sigma}_y(\hat{\mathcal{P}}-\hat{\mathcal{V}})^{-1}\right.\\ 
\left.\left\{\Delta\hat{N}_{\omega,q}-\Delta\hat{\mathcal{P}}(
\hat{\mathcal{P}}-\hat{\mathcal{V}})^{-1}\hat{N}_{\omega,q}-
\hat{N}_{\omega,q}(\hat{\mathcal{P}}^*-\hat{\mathcal{V}})^{-1}
\Delta\hat{\mathcal{P}}^*\right\}\right].
\end{split}
\end{equation}
This equation constitutes the essential result of this section. It relates the driving electric field to the polarization operators of the individual layers and the correlators of Langevin fluxes in them. The latter are determined by nonequilibrium distribution functions, which are in turn determined by the driving electric field via the system of equations (\ref{Poisson-phi})--(\ref{KE-average}). Further calculations of drag require explicit forms of the nonequilibrium corrections to the polarization function and the correlator of the Langevin fluxes. Determination of the nonequilibrium distribution function in the general situation is a very difficult problem. The situation simplifies dramatically when the rate of momentum transfer between  the layers due to drag is smaller than the equilibration rate of the electron liquid due to intralayer electron-electron scattering. This is always the case when the layers are sufficiently far apart. In this situation, the nonequilibrium state of the system may be characterized by two hydrodynamic velocities in each layer, and deviation from true thermal equilibrium is characterized by the difference between the layer velocities.

\section{Drag at fast intralayer equilibration}\label{sec:boost}

Drag resistivity may be characterized by the relaxation rate $1/\tau_D =\rho_D ne^2/m$ of the staggered momentum.~\cite{Smith,MacDonald} At large interlayer separations, this rate is significantly smaller than the intralayer equilibration rate of the electron fluid, $1/\tau$,
$\tau^{-1} \gg \rho_D ne^2/m$. Under this condition, the electron distribution function in each layer is well approximated by the equilibrium distribution with the drift velocity $\pm\mathbf{u}/2$ corresponding to the current density in the layer. In this case, we can obtain the nonequilibrium parts of various quantities by applying Galilean boost, namely
\begin{equation}
\Delta\hat{P}_{\omega,q}=\frac{\mathbf{qu}}{2}\hat{\sigma}_z\partial_\omega P_{\omega,q},\quad \Delta\hat{B}_{\omega,q}=\frac{\mathbf{qu}}{2}\hat{\sigma}_z\partial_\omega B_{\omega},
\end{equation}
where $\hat{B}_{\omega,q}$ is the matrix distribution function of collective bosonic excitations, which at equilibrium is given by $B_\omega=\coth(\omega/2T)$. Since by the fluctuation-dissipation theorem
\begin{equation}
\hat{N}_{\omega,q}=
-\frac{i}{2}\nu\left[\hat{P}_{\omega,q}-\hat{P}_{-\omega,-q}\right]\hat{B}_{\omega,q},
\end{equation}
we readily find in the notations of Eq.~(\ref{P})
\begin{equation}
\Delta\hat{N}_{\omega,q}=\frac{i\nu}{2}\left[(\Delta\hat{\mathcal{P}}-\Delta\hat{\mathcal{P}}^*)B+(\hat{\mathcal{P}}-\hat{\mathcal{P}}^*)\Delta\hat{B}\right].
\end{equation}
Using this result in Eq.~(\ref{Long-Trace}), and after some matrix algebra, we can rewrite the trace as a sum of two contributions,
\begin{subequations}\label{D-Force-T1+T2}
\begin{equation}e\mathbf{E}^i\bar{n}^i=\frac{i\pi\nu e^2}{4\epsilon\kappa^2}\sum_{q,\omega}\frac{q^2\mathbf{e_q}}{\sinh(qd)}
[\mathbb{T}_1+\mathbb{T}_2],
\end{equation}
\begin{equation}
\mathbb{T}_1=B_\omega\mathrm{Tr}\left[\hat{\sigma}_y[\Delta(\hat{\mathcal{P}}-\hat{\mathcal{V}})^{-1}-
\Delta(\hat{\mathcal{P}}^*-\hat{\mathcal{V}})^{-1}]\right],
\end{equation}
\begin{equation}
\mathbb{T}_2\!=\!-\frac{\mathbf{qu}}{2}\partial_\omega B_\omega\mathrm{Tr}\!\left[\hat{\sigma}_y(\hat{\mathcal{P}}-\hat{\mathcal{V}})^{-1}(\hat{\mathcal{P}}-\hat{\mathcal{P}}^*)\hat{\sigma}_z(\hat{\mathcal{P}}^*-\hat{\mathcal{V}})^{-1}\!\right]\!.
\end{equation}
\end{subequations}
It is easy to see that the variations $\Delta(\hat{\mathcal{P}}-\hat{\mathcal{V}})^{-1}$ and $\Delta(\hat{\mathcal{P}}^*-\hat{\mathcal{V}})^{-1}$ contain only matrices $\hat{\sigma}_0$, $\hat{\sigma}_x$ and $\hat{\sigma}_z$. Therefore, the trace of their product with $\hat{\sigma}_y$ must vanish identically, $\mathbb{T}_1=0$. The second trace does not vanish because of the explicit presence of the $\hat{\sigma}_z$ matrix arising from the staggered boost in the layers. Indeed, after cyclic permutation of matrices under the trace, we observe that
\begin{eqnarray}
(\hat{\mathcal{P}}^*-\hat{\mathcal{V}})^{-1}\hat{\sigma}_y(\hat{\mathcal{P}}-\hat{\mathcal{V}})^{-1}=\nonumber\\
\frac{\hat{\sigma}_y(|\mathcal{P}|^2-\mathcal{V}^2)+i\hat{\sigma}_z\mathcal{V}(\mathcal{P}-\mathcal{P}^*)}{|\mathcal{P}-\mathcal{V}|^2|\mathcal{P}+\mathcal{V}|^2},
\end{eqnarray}
where in accordance with Eqs.~(\ref{P}) and (\ref{V}) we wrote $\hat{\mathcal{P}}=\mathcal{P}\hat{\sigma}_0$ and $\hat{\mathcal{V}}=\mathcal{V}\hat{\sigma}_x$ with $\mathcal{P}=\mathcal{V}e^{qd}-P$.
As a consequence of the above equality, we find for the nonvanishing contribution to the trace in the formula for the drag force
\begin{equation}\label{T2}
\mathbb{T}_2=\frac{i(\mathbf{qu})}{2T\sinh^2(\omega/2T)}
\frac{\mathcal{V}(\mathcal{P}-\mathcal{P}^*)^2}{|\mathcal{P}-\mathcal{V}|^2|\mathcal{P}+\mathcal{V}|^2}.
\end{equation}
At this point we are prepared to find drag resistivity. Since averaging over the fluctuations has been performed, all the quantities appearing in the expression for the drag force should be understood as being equilibrium averages. In particular, to simplify notations in what follows we relabel the time average density $\bar{n}\to n$ with the understanding that $n$ is the equilibrium density of carriers in quantum wells.

We collect all the factors in Eqs.~(\ref{D-Force-T1+T2}) and (\ref{T2}), notice that upon angular averaging over the momentum transfer $\langle\mathbf{e_q}(\mathbf{e_qu})\rangle_\mathbf{q}=\mathbf{u}/2$, and thus we obtain for the drag resistivity $\rho_D=\mathbf{E}/en\mathbf{u}$ the following result
\begin{equation}\label{drag-resistivity}
\frac{\rho_D}{\rho_Q}=\frac{1}{8\pi Tn^2}\sum_{q,\omega}\frac{q^2\mathcal{V}^2}{\sinh^2(\omega/2T)}\frac{(\mathrm{Im}\mathcal{P})^2}{|\mathcal{P}-\mathcal{V}|^2|\mathcal{P}+\mathcal{V}|^2}.
\end{equation}
This expression can be brought to a more familiar form
by observing that $\mathrm{Im}\mathcal{P}=-\mathrm{Im}P$, and consequently
\begin{eqnarray}\label{Denominator}
&&\frac{\mathcal{V}(\mathcal{P}-\mathcal{P}^*)^2}{|\mathcal{P}-\mathcal{V}|^2|\mathcal{P}+\mathcal{V}|^2}
=\frac{\kappa^3}{q^3}\frac{(1-e^{-2qd})e^{-qd}(P-P^*)^2}{|\mathcal{E}_+\mathcal{E}_-|^2}\nonumber\\
&&=-\frac{\kappa}{4q}e^{-qd}\mathrm{Im}\left(\frac{1}{\mathcal{E}_+}\right)
\mathrm{Im}\left(\frac{1}{\mathcal{E}_-}\right),
\end{eqnarray}
where in a usual way we introduced the dielectric functions corresponding to the symmetric and antisymmetric density modes
\begin{equation}
\mathcal{E}_\pm(\omega,q)=1-\frac{\kappa}{q}(1\pm e^{-qd})P_{\omega,q}.
\end{equation}
As a result, Eq.~(\ref{drag-resistivity}) can be equivalently rewritten as
\begin{equation}\label{drag-resistivity-epsilon}
\frac{\rho_D}{\rho_Q}=\frac{1}{256\pi Tn^2}\sum_{q,\omega}\frac{q^2e^{-qd}\mathrm{Im}(\mathcal{E}^{-1}_+)\mathrm{Im}(\mathcal{E}^{-1}_-)}{\sinh^2(\omega/2T)\sinh(qd)}.
\end{equation}

It is instructive to make several comments in regards to Eq.~(\ref{drag-resistivity}). (\textit{i}) If one neglects the plasmon resonances in the denominator of Eq.~(\ref{drag-resistivity}), then one finds that drag is simply proportional to $(\mathrm{Im}P)^2$, which is a standard formula that applies to the contribution of the particle-hole continuum. (\textit{ii}) In general, the full expression (\ref{Denominator}) gives a complete formula that accounts for plasma resonance of the two dielectric functions. As will be explained later, plasma modes govern drag resistivity in the collision-dominated transport regime. (\textit{iii}) In the case of non-identical layers, Eq.~(\ref{drag-resistivity}) should be generalized as follows: in the numerator, one should replace $\mathrm{Im}(\mathcal{P})^2\to\mathrm{Im}\mathcal{P}^\uparrow\mathrm{Im}\mathcal{P}^\downarrow$ and $\mathcal{V}^2\to\mathcal{V}^\uparrow\mathcal{V}^\downarrow$, while in the denominator, $n^2\to n^\uparrow n^\downarrow$ and $|\mathcal{P}-\mathcal{V}|^2|\mathcal{P}+\mathcal{V}|^2\to|\mathrm{Det}(\hat{\mathcal{P}}-\hat{\mathcal{V}})|^2$, where the determinant can no longer be simply factorized as a product of two independent terms for each of the layers. In our analysis we concentrate on the case of symmetric layers. Even though the analytical calculations are still doable for the general nonsymmetric case, the obtained results for the drag resistivity become quite cumbersome and do not bring any new physics insight concerning the temperature dependence of the transresistivity.

\section{Relaxation time approximation}\label{sec:tau_approximation}

When analyzing temperature dependence of the drag resistivity from Eq.~(\ref{drag-resistivity}) we must discuss analytical structure of the polarization function $P_{\omega,q}$ in $(\omega,q)$-plane. In most of the previous studies this function was calculated for the bare noninteracting limit with respect to intralayer electron-electron collisions. This approximation is only sufficient to describe the low-temperature collisionless regime where drag is dominated by the particle-hole continuum. The finite intralayer equilibration length $\ell$ changes $P_{\omega,q}$ in a significant way, which becomes quantitatively important for drag already at moderately high temperatures $T>v_F/d$. Technically, finite $\ell$ makes $\mathrm{Im}P_{\omega,q}\neq0$ at the high-frequency limit, $\omega>v_Fq$, where the conventional collisionless result the yields vanishing spectral weight of particle-hole excitations. Consequently, as a result of equilibration processes, plasmon poles lie within the high-frequency tail of the an particle-hole continuum of states and yield important contribution to drag resistivity. This mechanism becomes progressively stronger as temperature gets higher, with thus a shorter equilibration length, and eventually plasmon modes take over and dominate the Coulomb drag. These points will be explicitly illustrated by the subsequent calculation, while we turn our attention now to the derivation of $P_{\omega,q}$ at finite $\ell$.

We treat the intralayer collision integral in the relaxation time approximation
\begin{equation}\label{St}
\mathrm{St}\{f\}=-\frac{1}{\tau}(\delta f-f_h).
\end{equation}
We assume that $\tau$ is dominated by electron-electron interactions, so that $\tau^{-1}\propto T^2$.
Importantly, the Fermi-liquid-like quadratic temperature dependence of the relaxation time has been experimentally confirmed even in the regime of correlated electrons with thus strong interactions $r_s\gg1$.~\cite{Eisenstein-tau}  In Eq.~(\ref{St}) $f_h$ stands for the hydrodynamic part of the nonequilibrium distribution, which has the locally equilibrium form
\begin{equation}
f_h=-\left[\delta\mu+m(\mathbf{u\cdot v})+\frac{\delta T(\varepsilon-\mu)}{T}\right]\partial_\varepsilon f_0
\end{equation}
where $f_0=[e^{(\varepsilon-\mu)/T}+1]^{-1}$ is the Fermi distribution. The Lagrange multipliers $\delta\mu$, $\mathbf{u}$ and $\delta T$ have the meaning of the local change in, respectively, the chemical potential, hydrodynamic velocity, and temperature. Their values are found from the conditions that the collision integral (\ref{St}) conserves the particle number, momentum, and energy:
\begin{subequations}\label{constraints}
\begin{equation}
\int d\Gamma \delta f=-\delta\mu\int d\Gamma\partial_\varepsilon f_0=\nu\delta\mu,
\end{equation}
\begin{equation}
\int d\Gamma\mathbf{p}\delta f=-\int d\Gamma\mathbf{p}(\mathbf{p\cdot u})\partial_\varepsilon f_0=mn\mathbf{u},
\end{equation}
\begin{equation}
\int d\Gamma (\varepsilon-\mu)\delta f=-\frac{\delta T}{T}\int d\Gamma(\varepsilon-\mu)^2\partial_\varepsilon f_0=\frac{\pi^2}{3}\nu T\delta T,
\end{equation}
\end{subequations}
with $\nu=\partial_\mu n$ being density of states. With the collision integral in the form of Eq.~(\ref{St})
the solution for the linearized Boltzmann-Langevin kinetic equation (\ref{KE-linearized}) can be presented in the form
\begin{eqnarray}
&&\delta f_{\omega,q}=\frac{1}{1-i\omega\tau+i\tau\mathbf{qv}}\left(\tau\delta J_{\omega,q}-\partial_\varepsilon f_0\times
\frac{}{}\right.\nonumber\\
&&\left.\left[ie\tau\phi_{q,\omega}\mathbf{qv}+\delta\mu_{\omega,q}+m\mathbf{vu}_{\omega,q}+\frac{(\varepsilon-\mu)\delta T_{\omega,q}}{T}\right]\right)
\end{eqnarray}
where we suppressed the layer index-$\imath$ for brevity and used the Fourier representation. We substitute the above expression into the integral constraints of Eqs.~(\ref{constraints}). It is convenient to separate the
integration over the phase space $d\Gamma$ into the integration over the energy and the angular
averaging over the Fermi surface, $\langle \ldots \rangle$. The latter can be performed with the aid of the following relations
\begin{subequations}
\begin{eqnarray}\label{eq:D}
\hskip-.45cm
\!\!\left\langle\!\frac{1}{1-i\omega\tau+i\tau\mathbf{qv}}\!\right\rangle \!
&\!=\!&\frac{1}{D_{\omega,q}},\\
\hskip-.45cm
\!\!\left\langle\!\frac{\mathbf{v}}{1-i\omega\tau+i\tau\mathbf{qv}}\!
\right\rangle\!
&\!=\!&-\frac{i\mathbf{q}}{q^2\tau}\left(1-\frac{1-i\omega\tau}{D_{\omega,q}}
\right),\label{eq:v_average}\\
\hskip-.45cm
\!\!\left\langle\!\frac{v_iv_j}{1-i\omega\tau+i\tau\mathbf{qv}}\!\right\rangle
\!&\!=\!&\!\left( L_{\omega,q}\!-\!M_{\omega,q}\right) \frac{q_iq_j}{q^2}\!+\!M_{\omega,q}\delta_{ij}
,\label{eq:v_i_v_j}
\end{eqnarray}
\end{subequations}
where we introduced the following dimensionless quantities
\begin{subequations}
\begin{eqnarray}\label{eq:D_def}
  D_{\omega,q}&=&\sqrt{(1-i\omega\tau)^2+q^2v^2\tau^2},\\
  \label{eq:L_def}
  L_{\omega,q}&=&\frac{1-i\omega\tau}{q^2\tau^2}
  [1-(1-i\omega\tau)/D_{\omega,q}]\\
  \label{eq:M_def}
  M_{\omega,q}&=&\frac{1}{q^2\tau^2}(D_{\omega,q}-1+i\omega\tau).
\end{eqnarray}
\end{subequations}
When performing integration in $d\Gamma$ over the absolute value of $v$ we make use of the advantage that in the degenerate electron liquid all integrals are dominated by momenta near the Fermi surface. As a result, we obtain a closed system of linear algebraic equations for the Fourier components of Lagrange multipliers
\begin{eqnarray}
&&\nu\left(1-\frac{1}{\mathcal{D}_{\omega,q}}\right)\delta\mu_{\omega,q}+i\nu\left(1-\frac{1-i\omega\tau}{\mathcal{D}_{\omega,q}}\right)\frac{m\mathbf{qu}_{\omega,q}}{q^2\tau}=\nonumber\\
&&e\nu\phi_{\omega,q}
\left(1-\frac{1-i\omega\tau}{\mathcal{D}_{\omega,q}}\right)+\int\frac{\tau\delta J_{\omega,q}d\Gamma}{1-i\omega\tau+i\tau\mathbf{qv}},\\
&&\nu\left(1-\frac{1-i\omega\tau}{\mathcal{D}_{\omega,q}}\right)\delta\mu_{\omega,q}-i\tau(n-\nu m\mathcal{L}_{\omega,q})\mathbf{qu}_{\omega,q}=\nonumber\\
&&e\nu\tau^2q^2\phi_{\omega,q}\mathcal{L}_{\omega,q}-\int\frac{i\tau^2\mathbf{vq}\delta J_{\omega,q}d\Gamma}{1-i\omega\tau+i\tau\mathbf{qv}},\\
&&\frac{\pi^2}{3}\nu T\left(1-\frac{1}{\mathcal{D}_{\omega,q}}\right)\delta T_{\omega,q}=\int\frac{\tau(\varepsilon-\mu)\delta J_{\omega,q}d\Gamma}{1-i\omega\tau+i\tau\mathbf{qv}},
\end{eqnarray}
where $\mathcal{D}_{\omega,q}$ and $\mathcal{L}_{\omega,q}$ are obtained from $D_{\omega,q}$ and $L_{\omega,q}$ with the replacement of $v$ by the Fermi velocity $v\to v_F$. It is interesting to observe that fluctuations of temperature decouple from the fluctuations of the density and drift velocity. Since our goal is to get the polarization function for the density fluctuations, we only need the first two of the above equations. Notice that the latter are also consistent with the continuity equation $-i\omega\delta n_{\omega,q}+in\mathbf{qu}_{\omega,q}=0$. By excluding $\mathbf{u}_{\omega,q}$ in favor of $\delta n_{\omega,q}=\nu\delta\mu_{\omega,q}$ from the above system, we can cast the equation for the density fluctuations in the form of Eq.~(\ref{Phi-Pi}) and consequently read off the polarization operator
\begin{equation}\label{Pi}
P_{\omega,q}\!=\!-\frac{\left(\omega+\frac{i}{\tau}\right)-\sqrt{\left(\omega+\frac{i}{\tau}\right)^2-q^2v^2_F}}{\left[1+\frac{2i\omega}{q^2v^2_F\tau}\right]\!\!\left[\big(\omega+\frac{i}{\tau}\big)-\sqrt{\big(\omega+\frac{i}{\tau}\big)^2-q^2v^2_F}\right]-\omega}.
\end{equation}
This important result together with Eq.~(\ref{drag-resistivity}) enables us to study the temperature dependence of the drag resistivity in various transport regimes, which is our immediate goal.

\section{Drag resistivity}\label{Sec-Results}

We have determined that there are three distinct temperature scales, and consequently four separate regions, where Coulomb drag resistivity exhibits qualitatively different behavior. At lowest temperatures, the equilibration length is very large, $\ell=v_F\tau\gg d$, electron kinetics is essentially collisionless, and drag is dominated by the particle-hole excitations with characteristic wave vectors $q\sim d^{-1}$ and frequencies $\omega\sim T$. Since the spectral edge of the particle-hole excitations is set by the line $\omega=v_Fq$, this introduces the characteristic crossover temperature $T_d\sim E_F/k_Fd$ [see Eq.~(\ref{eq:scales})].

As temperature increases, the equilibration length gets shorter and may become comparable to the interlayer spacing, $\ell\sim d$, which happens at the characteristic temperature
\begin{equation}\label{Tc}
T_c\sim E_F\sqrt{k_F/\kappa^2d}.
\end{equation}
This scale signifies the beginning of the collision-dominated transport regime. The remaining scale is deduced from the condition when the equilibration rate $\tau^{-1}$ becomes comparable to the energy scale of plasmon modes, $\omega_{pl}\sim v_F\sqrt{\kappa q}$, at characteristic for the drag wave vector $q\sim d^{-1}$. This yields the temperature scale
\begin{equation}\label{Th}
T_h\sim E_F\sqrt{k_F/\kappa}\sqrt[4]{1/\kappa d}
\end{equation}
above which hydrodynamic transport regime sets in. Our goal is to describe the crossover in the temperature dependence of drag resistivity throughout the entire range of hierarchical scales, $T_d<T_c<T_h$. Before we proceed with task it will be convenient to introduce several dimensionless variables and parameters. We will be measuring momenta $q$ and frequencies $\omega$ in units $d^{-1}$ and $T_d=v_F/d$ respectively, thus introducing $x=qd$ and $w=\omega/T_d$. We also introduce dimensionless equilibration length $l=v_F\tau/d$, and dimensionless frequencies $w^2_{s(a)}=\frac{\kappa d}{2}x(1\pm e^{-x})$ for symmetric (antisymmetric) plasmon modes. In these notations equation (\ref{drag-resistivity}) can be rewritten in a manifestly dimensionless form,
\begin{eqnarray}\label{rho}
\frac{\rho_D}{\rho_Q}&=&\frac{T_d}{32\pi T}\frac{1}{(nd^2)^2(\kappa d)^2}\sum_{x,w}\frac{x^4}{\sinh^2(x)\sinh^2(wT_d/2T)}\nonumber\\
&&\times\frac{\mathrm{Im}(P^{-1}_{w,x})^{2}}{\left|\frac{x^2}{2w^2_s}P^{-1}_{w,x}-1\right|^2
\left|\frac{x^2}{2w^2_a}P^{-1}_{w,x}-1\right|^2}.
\end{eqnarray}
We need to explore now various asymptotic limits of this formula.

\subsection{Collisionless regime $T<T_c$}

In the low temperature limit drag resistivity is dominated by the low-energy excitations since plasmon poles are much higher in energy. Indeed, for typical $x\lesssim1$ we have $w_{s(a)}\gg1$ while integral in Eq.~(\ref{rho}) is dominated by frequencies $w\ll w_{s(a)}$, so that we can replace the denominator in the second line of Eq.~(\ref{rho}) to unity  and obtain a simplified expression
\begin{equation}\label{rho-ph}
\frac{\rho^{ph}_D}{\rho_Q}=\frac{T_d}{32\pi T}\frac{1}{(nd^2)^2(\kappa d)^2}\sum_{x,w}\frac{x^4(\mathrm{Im}P^{-1}_{w,x})^{2}}{\sinh^2(x)\sinh^2(wT_d/2T)}.
\end{equation}
In the current notation with the superscript-$ph$ in $\rho^{ph}_D$ we want to emphasize that this contribution originates from particle-hole modes to distinguish it from contributions due to plasmons $\rho^{pl}_D$ that will be discussed below.

For $T\ll T_c$ equilibration length is $l\gg1$, and polarization operator in Eq.~(\ref{rho-ph}) can be taken in the main approximation. From Eq.~(\ref{Pi}) we find in the dimensionless variables
\begin{equation}\label{Pi0}
P_{w,x}=1-\frac{w}{\sqrt{w^2-x^2}}.
\end{equation}
For this temperature regime, keeping correction terms to $P_{w,x}$ in powers of $1/l$ will result only in subleading corrections to drag resistivity in Eq.~(\ref{rho-ph}) in a small parameter $T/E_F\ll1$. At the lowest temperatures, $T<T_d$, frequency integration in Eq.~(\ref{rho-ph}) is dominated by the range where $w\ll x$, so that it is sufficient to take $\mathrm{Im}(P^{-1}_{w,x})\approx w/x$, and thus we obtain
\begin{equation}\label{rho-ph-1}
\frac{\rho^{ph}_D}{\rho_Q}=\frac{\pi\zeta(3)}{16}\frac{1}{(k_Fd)^2(\kappa d)^2}\frac{T^2}{E^2_F},
\end{equation}
which is a well-known result.~\cite{Gramila,Smith,MacDonald,Kamenev,Flensberg}

At the higher temperatures, $T>T_d$, we can approximate $\sinh^2(wT_d/2T)\approx(wT_d/2T)^2$, however we need now the full expression for $P_{w,x}$ from Eq.~(\ref{Pi0}) since the leading contribution to integrals comes from $w\sim x$. As a result, drag resistivity crosses over to the linear temperature dependence
\begin{equation}\label{rho-ph-2}
\frac{\rho^{ph}_D}{\rho_Q}=\frac{\pi^3}{360}\frac{1}{(k_Fd)^3(\kappa d)^2}\frac{T}{E_F}.
\end{equation}
It is worthwhile emphasizing that the dependence on the interlayer separation changes as well from being proportional $d^{-4}$ to $d^{-5}$.

To account for the contribution from plasmon resonances to drag resistivity in this temperature regime, we need a more accurate expression for the polarization function in the high frequency limit. Expanding Eq.~(\ref{Pi}) under the assumption that $x/(w+i/l)\ll1$ one finds
\begin{equation}\label{Pi-high-frequency}
P^{-1}_{w,x}=\frac{2w^2}{x^2}-\left(1+\frac{w}{2(w+i/l)}\right).
\end{equation}
This expression allows one to approximate resonant denominators in Eq.~(\ref{rho}) by plasmon poles
\begin{equation}\label{plasmon-poles}
\left|\frac{x^2}{2w^2_\alpha}P^{-1}_{w,x}-1\right|^{-2}\!\!\approx\frac{w^4_\alpha}{[(w-w_\alpha)^2+\gamma^2_\alpha][(w+w_\alpha)^2+\gamma^2_\alpha]}
\end{equation}
where we have introduced plasmon damping rates
\begin{equation}\label{eq:plasmon-damping-rate}
\gamma_\alpha=\frac{x^2}{8}\frac{l}{1+(w_\alpha l)^2},\quad \alpha=s,a.
\end{equation}
The expression (\ref{eq:plasmon-damping-rate}) is based on the relaxation-time approximation and describes the plasmon attenuation rate in the entire crossover between the hydrodynamic ($w_\alpha l \ll 1$) and collisionless ($w_\alpha l\gg 1$) regimes. In the collisionless regime, it is consistent with the results of the microscopic treatment of plasmon attenuation rates in Ref.~\onlinecite{Mishchenko}.

For $T\ll T_c$ the plasmon contribution to $\rho_D$ in Eq.~(\ref{rho}) is dominated by the pole at the frequency of the antisymmetric plasmon. Using Eqs.~(\ref{Pi-high-frequency}) and (\ref{plasmon-poles}) in Eq.~(\ref{rho})
\begin{eqnarray}
\frac{\rho^{pl}_D}{\rho_Q}\!&\!=\!&\!\frac{T_d}{128\pi T}\frac{1}{(nd^2)^2(\kappa d)^2}\sum_{x,w}\frac{x^4}{\sinh^2(x)\sinh^2(wT_d/2T)}\nonumber\\
&\!&\!\frac{(wl)^2}{[1+(wl)^2]^2}\!\frac{w^4_a}{[(w-w_a)^2+\gamma^2_a][(w+w_a)^2+\gamma^2_a]}.
\end{eqnarray}
Integration over the frequency here is straightforward since poles near $\pm w_a$ are narrow and each Lorentzian function is effectively a delta-function with the weight $\pi/\gamma_a$. The remaining $x$-integration is dominated by the range where $x\lesssim 1$ where one can approximate $w^2_a=\kappa dx^2$ and find as a result
\begin{equation}\label{rho-pl-1}
\frac{\rho^{pl}_D}{\rho_Q}=\frac{3\zeta(3)}{16}\frac{1}{k^2_F\kappa d^3}\left(\frac{T}{E_F}\right)^5.
\end{equation}
This contribution is obviously smaller than that given by Eq.~(\ref{rho-ph-2}) in the same range of temperatures. We considered this term here in such details not only for completeness of our analysis. It will be shown in the next subsection that Eq.~(\ref{rho-pl-1}) will cross over into a different power-law behavior of drag resistivity that will overcome the respective contribution of the particle-hole continuum.

\subsection{Collision-dominated regime $T>T_c$}

We proceed to consider the intermediate and high temperature intervals. In this case the particle-hole continuum and the plasmon contributions are described by well separated peaks in the integrand of Eq.~(\ref{drag-resistivity}) and can be considered separately.

The particle-hole contribution is dominated by the frequency range where $w>x$ so that we need to use  Eq.~(\ref{Pi-high-frequency}) for the polarization operator. Furthermore, we may also approximate $\sinh(wT_d/2T)\approx wT_d/2T$ and set plasmon denominators in Eq.~(\ref{drag-resistivity}) to unity, which thus yields
\begin{equation}
\frac{\rho^{ph}_D}{\rho_Q}=\frac{T}{32\pi T_d}\frac{1}{(nd^2)^2(\kappa d)^2}\sum_{w,x}\frac{x^4}{\sinh^2(x)}\frac{l^2}{[1+(wl)^2]^2}.
\end{equation}
The remaining momentum and frequency integrations are elementary here, and one finds
\begin{equation}\label{rho-ph-3}
\frac{\rho^{ph}_{D}}{\rho_Q}= \frac{3\zeta(3)}{16}\frac{1}{(k_Fd)^2(\kappa d)^4}\frac{E_F}{T}.
\end{equation}
It is easy to check that at $T\sim T_c$ this result matches with Eq.~(\ref{rho-ph-2}), and at the same time it shows that the contribution of the particle-hole excitations to the drag resistivity becomes a decreasing function of temperature in the collision-dominated regime.

The situation with plasmon contributions is different.  Accounting for both symmetric and antisymmetric plasmon resonances, and using Eqs.~(\ref{Pi-high-frequency}) and (\ref{plasmon-poles}), we obtain from Eq.~(\ref{drag-resistivity})
\begin{eqnarray}
\frac{\rho^{pl}_D}{\rho_D}&=&\frac{T}{32\pi T_d}\frac{1}{(nd^2)^2(\kappa d)^2}\sum_{w,x}\frac{x^2}{\sinh^2(x)}\frac{l^2}{[1+(wl)^2]^2}\nonumber\\
&&\times\prod_{\alpha=a,s} \frac{w^4_\alpha}{[(w-w_\alpha)^2+\gamma^2_\alpha][(w+w_\alpha)^2+\gamma^2_\alpha]}.
\end{eqnarray}
Frequency integration in this formula can be done exactly, however it leads to an extremely cumbersome expression. A much more useful result can be obtained by exploring the following simplifying observation. The relevant wave numbers for the integrand function in the above expression are logarithmically large, $x\sim\ln(\kappa d/l^2)$. In this case, the frequencies of the symmetric and antisymmetric plasmons nearly coincide, and consequently their dumping rates become almost identical. Thus one can justify replacing $\gamma_\alpha\to\gamma=(x^2l/8)[1+\kappa dl^2x/2]^{-1}$ and also $l^2[1+(wl)^2]^{-2}\to(8\gamma/x^2)^2$. We integrate over $w$ by leading poles and make use of the following algebraic identities: $(w_sw_a)^2=\frac{1}{2}(\kappa dx)^2e^{-x}\sinh(x)$, $(w_s\pm w_a)^2=\kappa dx(1\pm\sqrt{1-e^{-2x}})$, $w^2_a+w^2_s=\kappa dx$, which leads us to the following intermediate result:
\begin{eqnarray}
\frac{\rho^{pl}_D}{\rho_D}&=&\frac{T}{32\pi T_d}\frac{l}{(nd^2)^2(\kappa d)}\sum_x\frac{x^3e^x}{\sinh(x)}\frac{1}{1+\kappa dl^2x/2}\nonumber\\
&&\times\left[1+\frac{x^3e^{2x}l^2(1+\sqrt{1-e^{-2x}})}{64\kappa d(1+\kappa dl^2x/2)^2}\right]^{-1}.
\end{eqnarray}
It is useful to observe at this point that $\kappa dl^2\sim (T_h/T)^4$ so that the remaining integral can be estimated for two asymptotic regions. At the range of intermediate temperatures, $T_c\ll T\ll T_h$, we estimate
\begin{equation}\label{rho-pl-2}
\frac{\rho^{pl}_D}{\rho_Q}\sim\frac{1}{(k_Fd)^4}\frac{T^3}{E^3_F},
\end{equation}
whereas in the hydrodynamic regime of high temperatures $T\gg T_h$ we estimate with logarithmic accuracy
\begin{equation}\label{rho-pl-3}
\frac{\rho^{pl}_D}{\rho_Q}\sim\frac{1}{(k_Fd)^2(\kappa d)^3}\frac{E_F}{T}
\end{equation}
in agreement with Ref.~\onlinecite{Hydro-drag} [see the comment in Ref.~\onlinecite{Remark}]. We can check that the low-temperature plasmon contribution (\ref{rho-pl-1}) matches with Eq.~(\ref{rho-pl-2}) at the expected crossover scale, $T\sim T_c$, while Eqs.~(\ref{rho-pl-2}) and (\ref{rho-pl-3}) are of the same order at $T\sim T_h$ where drag resistivity reaches its absolute maximum. It is instructive to compare particle-hole [Eq.~(\ref{rho-ph-3})] and plasmon [Eq.~(\ref{rho-pl-2})] contributions to drag resistivity in the collision-dominated regime. In particular, at $T\sim T_h$ one easily finds that plasmons dominate by a parametrically large factor $\rho^{pl}_D/\rho^{ph}_D\sim\kappa d\gg1$.

\begin{figure}
  \includegraphics[width=9cm]{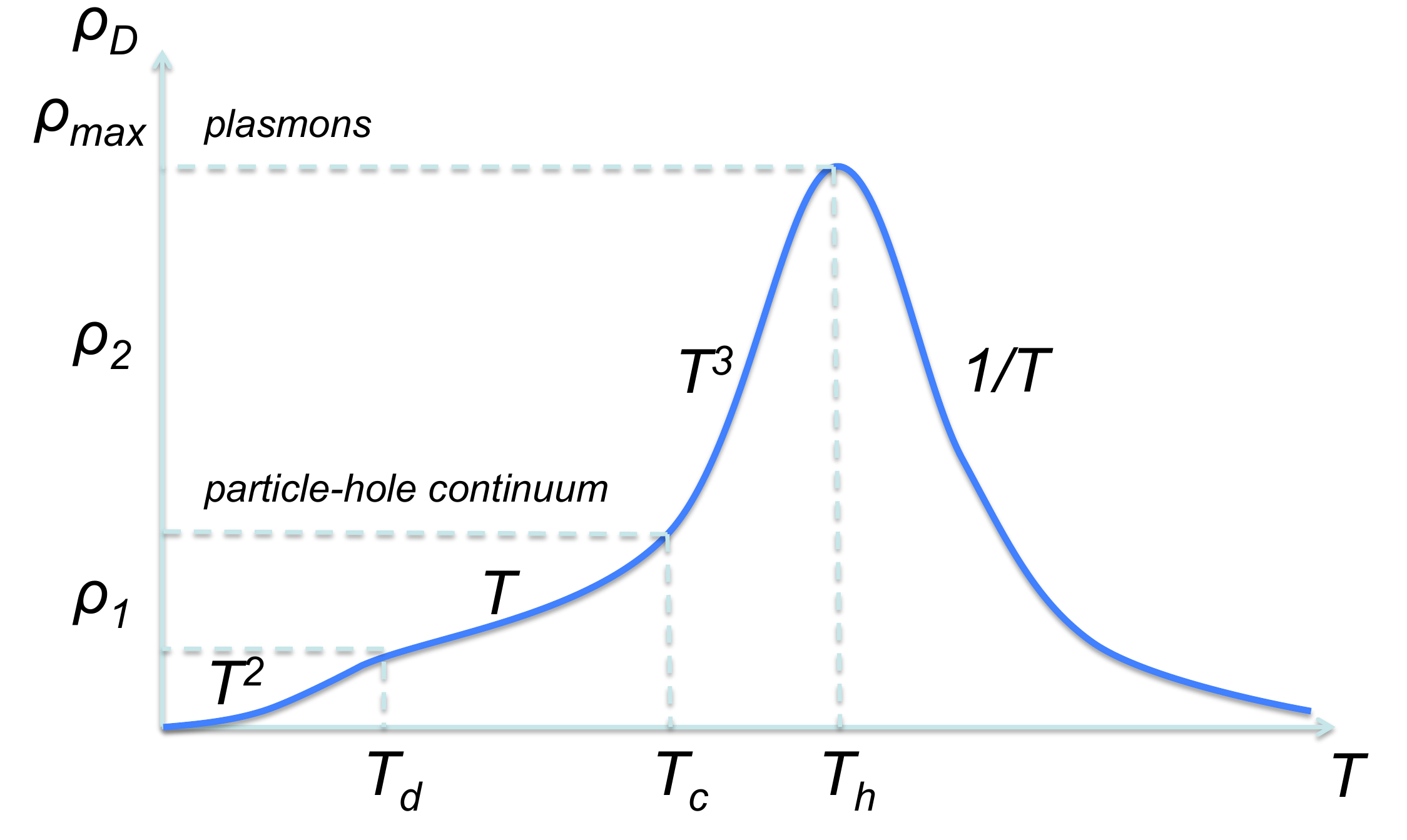}\\
  \caption{Schematic illustration of the temperature dependence of drag resistivity in various regimes. The four intervals from low to high temperatures are described by Eqs.~(\ref{rho-ph-1}), (\ref{rho-ph-2}), (\ref{rho-pl-2}), and (\ref{rho-pl-3}), respectively. At the crossover scales $T_d$ and $T_c$, the estimated magnitude of drag resistivity is $\rho_1/\rho_Q\sim1/(k_Fd)^4(\kappa d)^2$ and $\rho_2/\rho_Q\sim\sqrt{k_F/\kappa^2d}/(k_Fd)^3(\kappa d)^2$, while the maximum value of drag, $\rho_{max}$ is estimated in Eq.~(\ref{rho-max}). }\label{Fig}
\end{figure}

\section{Discussion}\label{Sec-Discussion}

We have developed a general computational scheme to describe nonlocal transport in interactively coupled double-layer systems. As an alternative route to existing formulations, our approach is based on the stochastic Boltzmann-Langevin kinetic theory. Using this approach, we reproduced the known results for the drag resistivity of clean double layers in the low-temperature regime, and we studied the previously unexplored collision-dominated transport regimes. Our main analytical result is Eq.~(\ref{drag-resistivity-epsilon}) for the drag resistivity, which is expressed in terms of the momentum- and frequency-dependent dielectric function of individual layers. This expression has some parallels with the Lifshitz theory of van der Waals forces.~\cite{Lifshitz} It holds as long as the intralayer equilibration rate is faster than the interlayer momentum relaxation rate. Note that this assumption is valid at large interlayer distances, $k_Fd \gg 1$.
Under these conditions, the electron distribution in the driven system is given by the locally equilibrium distribution function in each layer and is characterized by the corresponding drift velocity. The correlators of Langevin forces driving the density fluctuations can then be extracted from the fluctuation-dissipation theorem. Under more complicated nonequilibrium conditions, drag should be computed from the more general Eq.~(\ref{Long-Trace}) while the distribution function and Langevin forces should be found by solving the corresponding kinetic problem. It is worth stressing that the characteristic time scale for electron-electron scattering $\tau_{ee}$ can be significantly shorter than that for the impurity, $\tau_{im}$, or phonon, $\tau_{ph}$, scattering. For example, for relatively clean GaAs quantum wells at the doping level $\sim 10^{11}$cm$^{-2}$ and at temperatures $\sim 10$K electron-electron scattering time is estimated to be $\tau_{ee}\sim 10^{-11}$s whereas $\tau_{im}$ and $\tau_{ph}$ are typically of order the order of $\sim 10^{-9}$s. The relation $\tau_{ee}\ll\{\tau_{im},\tau_{ph}\}$ justifies applicability of our result for drag resistivity (\ref{drag-resistivity-epsilon}).

The general expression (\ref{drag-resistivity-epsilon}) applies not only to Fermi liquids but also to classical and semi-quantum strongly correlated liquids. To make further progress we focused on the Fermi-liquid regime. To obtain concrete expressions for the drag resistivity we treated electron-electron collisions in the relaxation time approximation. As a result we expressed the drag resistivity in terms of the single relaxation rate, $1/\tau_{ee}\propto T^2/E_F$. We found that there are four distinct temperature regimes in which drag exhibits a qualitatively different behavior and is governed by different types of density fluctuations -- quasiparticles and plasmons. The main findings of our work are summarized in Fig.~(\ref{Fig}), which schematically illustrates the temperature dependences of drag resistivity in the various transport regimes.

At low temperatures, $T<T_c$ [see Eq.~(\ref{eq:scales})], drag is dominated by the continuum of quasiparticle excitations with collisionless dynamics. Within that temperature domain, $\rho_D$ crosses over from being quadratic in $T$ below $T_d$ to a linear dependence in $T$ above $T_d$. Analytical expressions for these regimes are given by Eqs.~(\ref{rho-ph-1}) and (\ref{rho-ph-2}), respectively. Interestingly, it has been known from the early measurements that $\rho_D/T^2$ ceases to be a monotonic function of temperature already at low temperatures. Indeed, it has been pointed out in Ref.~\onlinecite{Gramila} that the ratio $\rho_D/T^2$ displays a noticeable falloff above a certain crossover temperature. Furthermore, it has been pointed out that the ratio $d^4\rho_D/T^2$ appreciably depends on $d$ above that temperature, and that the crossover scale itself shifts to lower temperatures for the larger interlayer separations. All these features are accounted for by our results for  particle-hole contributions to drag with the identification that expected crossover takes place at $T\sim E_F/(k_Fd)$.

The collision-dominated regime for quasiparticle dynamics sets in at $T>T_c$. In this regime, Coulomb drag is dominated by the plasmon contribution. The temperature dependence of this contribution is very sensitive to the plasmon attenuation rate. In the relaxation-time approximation, the latter is given by Eq.~(\ref{eq:plasmon-damping-rate}), which is consistent with the result of a microscopic calculation of Ref.~\onlinecite{Mishchenko}. It is worth noting that in a wide temperature interval, $T_c<T<T_h$ [see Eqs.~(\ref{Tc}) and (\ref{Th})], the frequency of the plasmon modes contributing to drag, $\omega_{pl}$ in Eq.~(\ref{eq:scales}), still exceeds the rate of electron collisions, so that their hydrodynamic treatment is inapplicable. In this intermediate regime, drag resistivity is $\propto T^3$, [see Eq.~(\ref{rho-pl-2})]. At $T> T_h$, the rate of equilibration exceeds $\omega_{pl}$. In this regime, drag can be treated using the hydrodynamic approach~\cite{Hydro-drag} and follows the $1/T$ temperature dependence [see Eq.~(\ref{rho-pl-3})]. In the entire range, drag may be represented by the interpolation formula
\begin{equation}
\frac{\rho_D}{\rho_Q}\sim \frac{1}{c_1(k_Fd)^2(\kappa d)^3T/E_F+c_2(k_Fd)^4E^3_F/T^3},
\end{equation}
with numerical coefficients of the order of unity, $c_1\sim c_2\sim1$. This expression allows us to estimate the maximum value of drag resistivity,
\begin{equation}\label{rho-max}
\frac{\rho^{max}_D}{\rho_Q}\sim\frac{\sqrt{\kappa/k_F}\sqrt[4]{\kappa d}}{(\kappa d)^3(k_Fd)^2}.
\end{equation}

We note that the measurements of Refs.~\onlinecite{Pillarisetty-1,Pillarisetty-2} in the low-density strongly correlated samples revealed that within the broad experimentally probed temperature regime, $\rho_D\propto T^{\alpha_T}/d^{\alpha_d}$, where power exponents vary in the range $1<\alpha_T<3$ and $2<\alpha_d<5$. Perhaps even more importantly, the observed magnitude of drag resistivity was one to two orders of magnitude larger than expected on the basis of a simple extrapolation of the small $r_s$ results. Such a manifestly nonquadratic temperature dependence of drag resistance combined with the unexpected magnitude of the effect was attributed in the literature to a possible non-Fermi-liquid behavior of strongly correlated liquids. In contrast, our conclusions are different. We have found that fast equilibration promotes a stronger drag effect, and at the same time it is responsible for the nonmonotonic temperature dependence of drag, so that most of the observations find a natural explanation within our theory. A very peculiar detail, which should be stressed once again, is that counterintuitively plasmons start to dominate the drag even at moderately high temperatures, and definitely at temperatures way below the plasmon resonances.

\section*{Acknowledgments}

We thank I. Aleiner, A. Kamenev, and B. Spivak for insightful discussions of the physical phenomena relevant to this work. A.L. acknowledges discussions with I.~Gornyi, B.~Narozhny, M.~Schuett, and M.~Titov concerning Refs.~[\onlinecite{Schutt,Titov,Narozhny-2}] and hydrodynamic theory of Coulomb drag.

W.C. acknowledge the support of NSFC under grant No. 11447602. Work at the University of Washington was supported by the U. S. Department of Energy Basic Energy Science Program under the award number DE-FG02-07ER46452. This work of A.L. was supported by NSF Grant No. DMR-1401908.


\begin{thebibliography}{99}
\bibitem{Remark}
In Ref.~(\onlinecite{Hydro-drag}), the plasmon contribution to drag in the hydrodynamic regime was expressed in terms of the viscosity of the electron liquid (up to a  logarithmic factor) as $\rho_D/\rho_Q\simeq \epsilon T\eta/e^2n^4d^5$. Assuming the Fermi-liquid regime and using the usual kinetic formula for viscosity, $\eta\sim mv_Fn\ell\sim nE^2_F/T^2$, one recovers our Eq.~(\ref{rho-pl-3}). Note also that our formula (\ref{rho-pl-3}) does not include an additional contribution to drag due to thermal modes discussed in Ref.~(\onlinecite{Hydro-drag}), since it is subleading in the hydrodynamic regime.
\end{thebibliography}
\end{document}